%% file: proceedings.tex
\documentclass{sigchi}

\usepackage{balance}  
\usepackage{graphics} 

\usepackage{txfonts}
\usepackage{times}    
\usepackage[pdftex]{hyperref}
\usepackage{color}
\usepackage{textcomp}
\usepackage{booktabs}
\usepackage{ccicons}
\usepackage{todonotes}
\usepackage{changepage}
\usepackage{amsmath,amssymb}
\usepackage{url}
\usepackage{array}

\makeatletter
\def\url@leostyle{%
  \@ifundefined{selectfont}{\def\UrlFont{\sf}}{\def\UrlFont{\small\bf\ttfamily}}}
\makeatother
\def\pprw{8.5in}
\def\pprh{11in}

\definecolor{linkColor}{RGB}{6,125,233}

\definecolor{jess}{RGB}{102,166,30}
\definecolor{jomara}{RGB}{240,128,128}
\definecolor{jack}{RGB}{217,95,2}

\hypersetup{%
  pdftitle={SIGCHI Conference Proceedings Format},
  pdfauthor={LaTeX},
  pdfkeywords={SIGCHI, proceedings, archival format},
  bookmarksnumbered,
  pdfstartview={FitH},
  colorlinks,
  citecolor=black,
  filecolor=black,
  linkcolor=black,
  urlcolor=linkColor,
  breaklinks=true,
}

\begin{document}

\title{Community Animation: Exploring a design space that leverages geosocial networking to increase community engagement}

\numberofauthors{3}
\author{%
  \alignauthor{Jomara Sandbulte\\
    \affaddr{Pennsylvania State University}\\
    \affaddr{University Park, PA USA}\\
    \email{jmb89@psu.edu}}\\
  \alignauthor{Jessica Kropczynski\\
    \affaddr{University of Cincinnati}\\
    \affaddr{Cincinnati, OH USA}\\
    \email{jess.kropczynski@uc.edu}}\\
  \alignauthor{John M. Carroll\\
    \affaddr{Pennsylvania State University}\\
    \affaddr{University Park, PA USA}\\
    \email{jmc56@psu.edu}}\\
}

\maketitle


\input{00_Abstract.tex}



\input{10_Introduction.tex}
\input{20_RelatedWork.tex}
\input{30_Methods.tex}
\input{40_Results.tex}
\input{50_Discussion.tex}
\input{60_Conclusion.tex}

\bibliographystyle{SIGCHI-Reference-Format}
\bibliography{Community2017}

\end{document}

%% file: 00_Abstract.tex
\section{Abstract}

This paper explores a design study of a smartphone enabled meet-up app meant to inspire engagement in community innovation. Community hubs such as co-working spaces, incubators, and maker spaces attract community members with diverse interests. This paper presents these spaces as a design opportunity for an application that helps host community-centered meet-ups in smart and connected communities. Our design study explores three scenarios of use, inspired by previous literature, for organizing meet-ups and compares them by surveying potential users. Based on the results of our survey, we propose several design implications and implement them in the Community Animator geosocial networking application, which identifies nearby individuals that are willing to chat or perform community-centered activities. We present the results of both our survey and our prototype, discuss our design goals, and provide design implications for civic-minded, geosocial networking applications. Our contribution in this work is the development process, proposed design of a mobile application to support community-centered meet-ups, and insights for future work.

%% file: 10_Introduction.tex
\section{Introduction}
Urban planners have recognized the growing number of co-working and innovation spaces that bring together groups with different interests and pose an opportunity to stimulate local economies and civic activities. A challenge in these spaces, is to leverage opportunities to engage multi-talented individuals in projects, activities, initiatives, and conversations that will enhance the community and local economy. The role of the Community Animator (CA) has emerged to address this challenge. CAs are an emerging role and job title within these innovation spaces \cite{coalition2015,linkedin2015,animation2010}. CAs participate in and manage communities by sharing information and building bridges between members \cite{coalition2015, enacademic2015,animation2010}. In these spaces, the CA acts as a party host who ``animates" a community by hosting introductions based on similar interests. Our study aims to explore this design space to provide this same ``animation'' to communities beyond innovation spaces and enhancing civic interactions more broadly. We leverage similar functionality to popular meet-up geosocial networking applications to animate communities around civic issues in community hubs and beyond. One challenge in designing for this space is selecting a facilitation mechanism to host introductions. Community engagement apps have developed platforms that inspired citizen interactions through a variety of methods, e.g. common interests \cite{Nguyen:2015:KSS:2702123.2702411,McCarthy:2009:SCT:1556460.1556493}, discussing local news \cite{han2014local}, or participating in tasks \cite{han2015s}. 

First, we describe three scenarios borrowed from previous literature in order to more clearly define the requirements of geosocial networking applications that engage citizens within a community. These scenarios are meant to illustrate how we can effectively inspire community-centered activities such as: discovering new resources in the community \cite{carroll2006social,carroll2014presence}; learning about others perspectives \cite{han2014local}; or filling interstitial time \cite{carroll2013co,dimmick2010news} that would otherwise be spent not engaging with one's community. Our research identifies appropriate topics to promote specific outcomes relevant to sense of community and community engagement, in order to guide the direction of how to promote civic interaction within a community. This work contributes an identification of a design opportunity and a requirements analysis through the testing of scenarios with potential users. Second, we propose and evaluate a design based on these findings. Our evaluation involved hosting an event where participants were guided to create a profile and asked to meet-up with matching profiles. After using the application, the participants then responded to a survey to describe their experience using the app. We present our findings and relate them to our design guidelines and considerations, and conclude with future work.

\subsection{Research Questions}
Technology that links instrumental aims with a community to create innovative solutions for the local community's benefit is difficult to design without understanding the value it can have for community members. Therefore, the overarching question guiding this research is: 

\textit{What are the design requirements for meet-up apps to automate community animation and strengthening local community engagement? }

We start by probing the perceptions of our design scenarios with potential users, which we later used to refine our design. Therefore, the research questions for the first phase of our investigations were: 

\textit{(RQ1) What value (if any) do potential users believe they will gain while using the app in various scenarios of use in their community?}

\textit{(RQ2) How do potential users believe Community Animator use can contribute to elements of engagement in their community and why?}

Upon answering these two research questions we developed and deployed a working prototype of our app to further investigate users’ perceptions and further test our design hypothesis of producing a geosocial networking application that facilitates community animation. Our research question for this phase is: 

\textit{(RQ3) How do potential users believe the app will contribute to their engagement in community innovation and why?}

%% file: 20_RelatedWork.tex
\section{Related Work}
Previous work has indicated that when it comes to developing community innovations, the collective vision of local businesses, organizations, educational institutions, and local government are often disconnected, even though they share a common goal for positive change \cite{carroll2006social,carroll2014presence}. McCarthy \cite{McCarthy:2009:SCT:1556460.1556493} affirmed community spaces (often referred to as third places \cite{oldenburg1989great}) benefit communities by enabling people to discuss, plan and execute ``potentially useful collective undertakings." A lack of these third spaces removes the social environment necessary to facilitate civic interactions. To avoid the increase of social isolation and loss of social connections, urban planners are putting more effort into creating new third places and reinvigorate metropolitan neighborhoods by breaking down social siloes \cite{thirdplacesbuilders}. For instance, the OUTBOX experiment \cite{outboxexperiment} provided an outside and wall-less vestibule equipped with Wi-Fi and seating, available for people to use throughout the day.
Web media such as \textit{Dodgeball.com, MeetUp.com, Foursquare,} and \textit{Facebook Check In} are examples of geosocial platforms that enable meeting others, which are experiencing a boom in popularity. These options work with user's location to offer community groups or to spread information about a local place. Dating applications such as \textit{Tinder} and \textit{OkCupid} are examples of geolocation sites increasingly used for social interaction by connecting individuals in person. However, these online dating applications differ from other types of social interaction as they enable only pairwise interactions, which limits their potential for community expansion.

Interactions around local place have long been theorized in community informatics literature. Carroll \cite{carroll2015community} describes the community informatics as a field concerned with the challenges and opportunities for citizens in an environment increasingly dominated by technology. Civic technology holds the potential to motivate local citizens to become active in community events \cite{gurstein2003effective,knightcivic2015}. Considerable efforts have been dedicated to the development of applications that integrate innovations and technology to facilitate community citizen engagement. Various designs have been developed to inspire meet-ups and engage citizens in their communities. Here we explore previous methods used to host introductions using mobile apps. 

\subsection{Interest-based engagement}
There are a variety of applications that have been developed to engage social communication and facilitate ``ice breaking" based on shared interests. For instance, the SoBot application \cite{Xu:2014:SFC:2556420.2556789} intends to promote a conversation and facilitate introducing two people. SoBot uses social data available online to create a profile and tries to determine the similarity of users' interests by using recommendation algorithms. The Community Collage (``CoCollage") proposed by McCarthy et al. \cite{McCarthy:2009:SCT:1556460.1556493} is a place-based social networking application which links online interests, activities, and physical presence to introduce people in coffeehouses.
In addition, there are also studies of systems made for wearable devices like \textit{Google Glass} \cite{Nguyen:2015:KSS:2702123.2702411} that are meant to ease conversation between strangers by recommending conversation topics for them to discover shared interests. There are also devices on the market that help users meet new people. \textit{Amico} bracelets \cite{amico2015} are wearable devices that allow users to create an interest-based profile that syncs with the wristband. When a user crosses the path of a person with a similar interest, the user receives a notification on the device. These are examples of workable interest-based methods of enhancing experience that may be appropriated for meet-ups based on civic interests. 

\subsection{News-based engagement}
Tools designed for engagement in local communities have worked to promote online community conversations surrounding local news and events \cite{carroll2006social,carroll2014grounding,carroll2014presence,Kavanaugh:2012:LNA:2307729.2307736,Kavanaugh:2013:ECP:2479724.2479750}. These projects collected data from local news, events, and calendar (e.g.\textit{ Local News Chatter, Arts Fest, CiVicinity, Outside.in}) and local heritage (e.g.\textit{Lost State College, Future State College}) to inform the local population about upcoming events and trending topics. Some of these tools have been effective in reinforcing local citizen awareness of local news, granting access to different perspectives, and enabling users to share their own interests and concerns \cite{carroll2014presence,Han:2014:SAM:2556420.2556824,Kavanaugh:2013:ECP:2479724.2479750}.

\subsection{Task-based activities}
The \textit{hOurWorld} mobile app is a design approach that promotes civic culture based on performing need-based tasks \cite{han2015s}. In this scenario however, interactions are based on hour exchanges where one party spends hours and another will earn hours that they can ``bank" for a later date. Although tasks are being performed in this instance, they are not necessarily performed together. 

%% file: 30_Methods.tex
\section{Methods}
Given the themes of interest-, news-, and task-based community engagement interventions present in previous literature, we leverage them in order to design three scenarios of use. Each scenario is meant to support meet-up criteria that facilitate engagement in smart and connected communities. Scenario-based design is a common tool for requirements analysis that aids design for specific interaction sequences \cite{Rosson:2001:UES:501581}. We are similarly interested in how likely a scenario of use identified in existing literature (compared to alternate scenarios of use) is to inspire community-centered activities. 

We began by designing three scenarios based on the three strategies that emerged from the research themes of previous research \cite{carroll2006social,carroll2014grounding,carroll2014presence,han2014local,han1916being,Kavanaugh:2012:LNA:2307729.2307736,Kavanaugh:2013:ECP:2479724.2479750,shih2015engaging,Xu:2014:SFC:2556420.2556789}  that illustrated how each could be used to promote engagement with other members of the local community in geosocial networking apps. Our survey investigates which scenario will be most useful in achieving the community-centered activities necessary for the hypothesized contribution. Each of the scenarios started with a user discovering the app for the first time, selecting profile items based on a particular scenario (e.g. to find people based on similar general interest categories, to find people to discuss local news stories with, or to find people that would like to collaborate on a common task in the local area), and then engaging in a meet-up based on the scenario. A summary of each of the three scenarios follows: 

\textit{Scenario: Shared interest in environment}\\
This scenario describes a person who is interested in sustainable agriculture. He shops at the farmer's market, chats with close friends and follows related news feeds on social media, but he is unaware of other forums to discuss this topic. Upon entering a local deli, he sees a sign about an app that uses GPS on smartphone to search for citizens nearby his location with similar local interests and hosts an introduction so they might carry out a face-to-face conversation. He downloads the application, selects ``Food and Agriculture" as one of his interest areas, indicates he is ``animated" and available for conversation. Soon after, he is introduced to someone also having lunch and indicated the same interest. They sit together while eating and each learn more about vendors and events they were previously unaware of. 

\textit{Scenario: News about local area}\\
This scenario describes a student who is interested in architecture and often follows the local newspaper on social media and read stories about local development. She is excited to see that high-rise student apartment buildings with a modern design are being proposed for development in the downtown. She sees the value of these opportunities for creating new residences within walking distance to campus and local businesses for maintaining a vibrant downtown and reducing rush hour traffic in an already congested area. Through the comments in the newspaper she sees that many community members do not feel the same way, and instead would prefer that new residencies were created away from the downtown, although many occupants would likely be students that will contribute to rush hour traffic near campus and the downtown. She learns of a GPS enabled app to search for citizens nearby her location who have read similar local news stories and hosts an introduction so they might carry out a face-to-face conversation. She downloads the application, selects ``New High-Rise Apartment Building" as a new story that she is eager to discuss and is introduced to a local resident that also interested in the same issue. The two learn from one another's perspective and reshape their own opinions of the proposed high rise.

\textit{Scenario: Volunteering as a shared task}\\
This scenario described someone who is interested in being more involved in volunteering since it has previously given him the chance to meet new people while doing something constructive for his community. He downloads an app while downtown at a community hub to search for nearby users that are ``animated" in a task that could be carried out as a group. He downloads the app and selects ``Volunteerism" and is then introduced to a person is in the same building and is involved with litter cleanup. The two meet-up and walk to the cleanup site with a group of students who share the same desire keep the community clean. While cleaning together they brainstorm more flash cleanup opportunities around the area.

\subsection{Scenarios Videos}
\begin{figure}[!ht]
\centering 
\includegraphics[width=9cm]{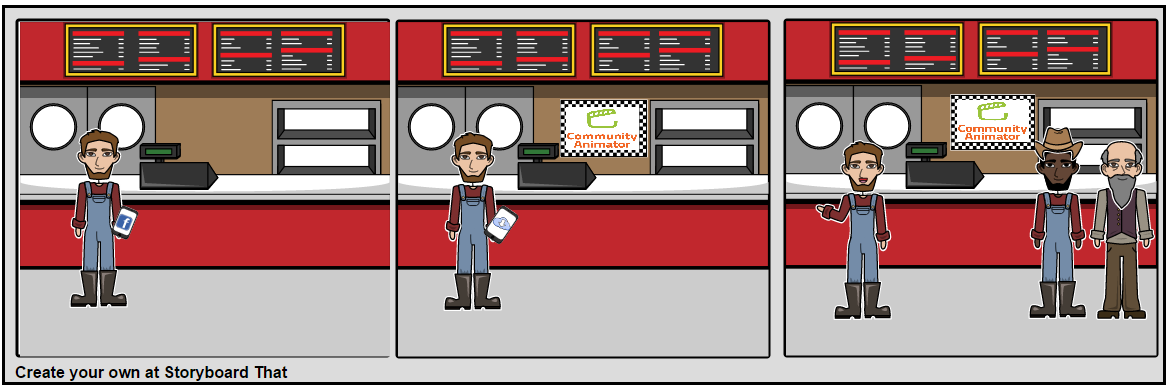}
\caption{Illustration of the scenario comic strip}
\label{figure:scenario}
\end{figure}

We created a video illustrating the scenarios through voice over narration of comic strips with closed captioned transcripts. Each video was read by the same presenter. The comic strips were each created with a similar aesthetic using comic strip characters generated by \textit{StoryboardThat.com} as illustrated in Figure \ref{figure:scenario} \cite{storyboard2014}. The videos were embedded with random order of display of scenario-based question blocks within a Qualtrics' survey.

\subsection{Survey Questions}


All survey questions asked the participants to indicate their level of agreement on a 10-point Likert scale, that varied from strongly disagree (value of 1) to strongly agree (value of 10) on what would be their perceived likelihood of downloading and using an app that supports the scenario. Scales to measure community-centered activities were adapted from questions used in  community collective efficacy scales \cite{carroll2003community,Carroll:2005:CEM:1054972.1054974} and have since been abbreviated and used in previous literature \cite{han2014local,han1916being,shih2015engaging} in this area (see Table \ref{table:survey}). 

In additional to questions related to their perception of the app, we also engaged participants in exploratory work of interest-, news-, and task-based meet-up categories that would be of interest to them for each scenario. For example, participants were asked, ``We have identified the following 15 interest-based categories that may inspire community members to start a conversation. Please check which categories you feel would be useful to hosting a meet-up like the one in the video and use the open-ended area that follows to help us brainstorm additional categories."

\subsection{Procedure}
 We sent recruitment emails  to local community organizations and published a note to a university-affiliated research bulletin.  The emails were shared widely and were made available for one month. Participants watched videos describing all three scenarios. After each video, participants were then asked whether they believed they would use the app to engage in various community activities or not. Our motivation was to understand participants willingness to download and use an app because some critical mass will be necessary for users to have any chance of meeting one another and facilitate such a scenario. 

\subsection{Participants}
We had a total of 109 participants that started the survey and 98 complete the survey. Of those, 64\%\ were female and 36\%\ male. Participants were between the age of 19-67 years old and the mean age was 32 years old. We utilized occupation categories by the US Census, and we found that the largest portion respondents were associated with higher education, which is not surprising for a college town, 27\%\ of the participants were undergraduate students, 13\%\ graduate students, and 14\%\ were in a profession associated with education. These demographics were similar to that of those engaged in the co-working and innovation spaces in the local community that are comprised of interns, volunteers, co-workers, employees, and members that participate in programs and services in these spaces.

%% file: 40_Results.tex
\section{Results of Survey}
Results of the willingness to download comparison are shown in Table \ref{table:demographics}. We conducted a one-way between subjects ANOVA to compare the effect of scenarios on willingness to download the app to discuss interests, discuss news, or collaborate in local tasks. There was a significant effect of the scenario type on willingness to download at the p \textless.05 level for the three scenarios [F(2,179)=3.74, p=.026].

\begin{table}[!ht]
\setlength\extrarowheight{3pt}
\centering
\caption{Participants perceived willingness to download and use app (scale from 1-10).}
\label{table:demographics}
\begin{tabular}{p{0.2\textwidth}|p{0.08\textwidth}|p{.06\textwidth}}
\hline
\textbf{Scenario of Use} & \textbf{Average} & \textbf{SD} \\ 
\hline
Interest-based meet-up & 5.62  & 2.62 \\ 
\hline
News-based meet-up & 5.69  & 2.56 \\ 
\hline
Task-based meet-up & 6.77 & 2.56  \\ 
\hline
\end{tabular}
\end{table}

When participants were asked which of the three scenarios were most useful to achieving the community-centered meet-ups, potential users indicated the task based scenario and ranked interest and news equally on a 10-point Likert scale (see table \ref{table:survey}). Although this is the case, when asked questions related to perceived community-centered activities from app use when all three scenarios were compared, local news items were rated the lowest on a 10-point Likert scale than other meet-up criteria (see table \ref{table:survey}). Although interest-based meet-ups rated slightly lower in terms of willingness to download, on average, participants rated shared interest higher than other scenarios when it comes to learning about others perspectives in the community and discovering new resources. On average, participants rated common tasks as likely to aid in meeting new people, discussing issues that they would not discuss otherwise, fill downtime that would otherwise not be spent engaging in the community. Two additional goals of our app is to encourage users to share basic information (such as name, profile photo and a location) about themselves while animated and to meet with someone in person rather than having a text-based conversation. When we asked potential users about each of these topics, there was not a significant difference between means, but on average, participants stated that they would be most likely to have a text-based conversation rather than meeting in person relating to shared interest, and would be most likely to disclose basic information about themselves while using the app to find others to engage in a common task.

\begin{table}[!ht]
\begin{flushleft}
\setlength\extrarowheight{4pt}
\centering
\small
\caption{Responses related to community-centered activities (scale from 1-10).}
\label{table:survey}
\begin{tabular}{m{7em}|m{1cm}m{.7cm}m{1cm}m{1cm}m{.9cm}}
\hline
\multicolumn{1}{c |}{\textbf{Survey Questions}}\\Would you use this app to... & \textbf{Shared Interest (Mean)} & \textbf{Local News (Mean)} & \textbf{Common Task (Mean)} & \textbf{F-ratio} & \textbf{p-value}\\ 
\hline
...have a text-based conversation (through a text  messages) rather than meet with someone in person? & 5.39 & 4.57 & 4.61 & 1.384 & .253\\ 
\hline
...meet new people? & 5.74 & 4.36 & 6.56 & 9.075 & .000 \\ 
\hline
...learn about others perspectives in your community? & 7.15 & 5.58 & 6.81 &  1.394 & .251 \\ 
\hline
...discover new resources in your community? & 7.15 & 5.58 & 6.81 & 6.070 & .003 \\ 
\hline
...discuss issues that I would NOT talk about otherwise? & 4.14 & 4.00 & 4.68 & .736 & .481 \\
\hline
...to fill downtime that might otherwise be spent not being engaged in my community? & 5.43 & 4.18 & 5.47 & 4.027 & .020 \\
\hline
...disclose basic information about myself while ``animated" (such as name, profile photo, and location)? & 5.55 & 4.78 & 5.77 & 1.764 & .174 \\
\hline
\end{tabular}
\end{flushleft}
\end{table}

\section{IMPACT ON THE DESIGN}
Our findings provide insight into the particular challenges in designing for community innovation and, perhaps even more importantly, insight into how to address the challenge of recruiting citizens to help address this challenge.
Table \ref{table:demographics} shows that users feel relatively indifferent to adopting a meet-up app based on interests or news stories, however, task based meet-ups seems to be a scenario that may inspire more downloads (RQ1).
Although this is the case, we were reluctant to abandon the interest-based scenario as respondents indicated that it holds potential for users to discover new resources. In contrast, participants rated news-based categories lowest in terms of its potential to host community-centered meet-ups.
Table \ref{table:survey} shows that users rated the instrumental advantage of discovering new resources higher than many community building community-centered activities. Our aim is to create opportunities for both instrumental and community-centered activities.
In designing the app, we found potential to emphasize tasks, but also include ``conversations" as a task category that may embed further interest-based areas if selected by a user (RQ2). 

Based on the survey results, we decided to go forward on designing and developing the Community Animator mobile app prototype. The survey results informed our decision to retain categories related to both interest-based and task-based categories for meet ups, and to eliminate the news- based scenario due to few responses indicating that its use would be used for the desired community centered activities. When constructing our design, we saw an opportunity to easily incorporate these two categories by asking participant to select from a list of tasks. One task that we included is conversation. Under conversation is the list of interest-based conversation topics. 

\begin{figure}[!ht]
\centering 
\includegraphics[width=9cm]{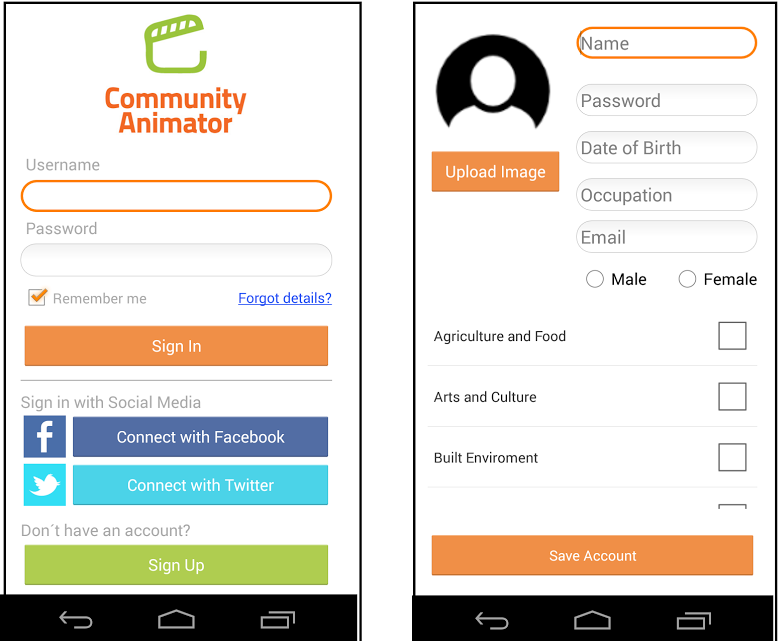}
\caption{Screenshots of Community Animator, a potential implementation of design.}
\label{figure:screenshots}
\end{figure}

When a member of the community downloads the app, it first prompts the user to agree to terms and conditions that includes permission to use their in-app activity for research purposes and provides the elements of informed consent that users must agree to in order to continue.
It then asks the community member to create a sign-in \textit{username} for the in-application account (Figure \ref{figure:screenshots}, left) or create an account using an existing social media account.
Next, the community member will be asked to share a name to be displayed to other community members using the app.
This field is populated by social media data (if used) but may be changed to a pseudonym by the community member.
Below, they can scroll and select through 33 civic tasks identified by survey participants.
By checking the boxes as they scroll, the community member identifies that this is an activity that they would like to participate in when ``animated" (Figure \ref{figure:screenshots}, right).
Conversations is one of the tasks that users may select while animated; if selected, the user will be shown a list of 26 interest-based areas that they can elect to engage in conversation about by indicating that they would like to meet-up with others with this conversation interest area.

\begin{figure}[!ht]
\centering 
\includegraphics[width=9cm]{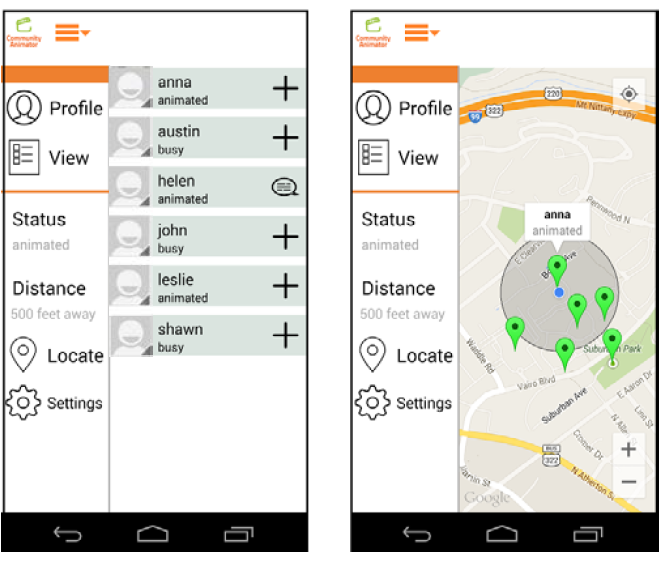}
\caption{Screenshots of Community Animator visualization options.}
\label{figure:screenshots1}
\end{figure}

Once an account is created and task categories have been selected, the community members' geolocation of their smartphone allows Community Animator to connect and coordinate with users in close proximity with matching civic categories as illustrated in Figure \ref{figure:screenshots1}. Users could select list view (Figure \ref{figure:screenshots1}, left) and map view (Figure \ref{figure:screenshots1}, right) to visualize animated participants close to his or her location.

\section{EVALUATION}
We conducted two separate evaluations of Community Animator app: a cognitive walk through survey with a group of potential users from a general audience to engage in a ``happening" and a focus group with members of a local community hub that also serves as a co-working space located in the local municipal building.

\subsection{Walk Through Event}
A test of the initial prototype took place at a staged happening. We recruited participants by posting fliers and inviting anyone who was interested in testing an app and meeting new people to meet at a campus caf\'e.
Participants were asked to register in advance for the event via an online survey, right before the event registered participants were emailed instructions about how to download the app to their mobile device.

On the day of the event, 14 participants joined in the large caf\'e where others were enjoying lunch, conversation, or work.
These others were not aware of the happening.
Participants were asked as they arrived to spread out into separate locations in the caf\'e and to create an account on the app and set their status to ``animated". We observed participants starting by spending a few minutes reading through interest and task categories and selecting categories. Then, they moved to the screen that showed matches in the area. Participants then selected another user from a list, and engaged in some in-system chat to coordinate a meet-up. We asked participants to simply enjoy conversations and the activity in order to best share an experience using the app. Afterwards, participants were requested to respond to a post-event online survey.

\subsection{Results of App Activity Data on Walk Through Event}

Based on survey results, our app allowed users to select meet-up items related to both tasks and interests.
Table \ref{table:interests} shows the number of interests versus the number of tasks that were selected by each user when creating an account. 

\begin{table}[!ht]
\setlength\extrarowheight{4pt}
\centering
\caption{Number of interests and tasks that each participant selected.}
\label{table:interests}
\begin{tabular}{p{0.1\textwidth}|p{0.07\textwidth}|p{.15\textwidth}|p{.05\textwidth}}
\hline
\textbf{Number of} & \textbf{Average} & \textbf{Range(Min, Max)} & \textbf{SD} \\ 
\hline
Interests & 4.583 & (3, 8) & 1.730 \\ 
\hline
Tasks & 3.667 & (1, 10) & 2.348  \\ 
\hline
\end{tabular}
\end{table}

By asking users to select tasks that they were interested in, our intention was to promote meet-ups with other attendees based on their profile preferences rather than pre-assigned categories, simply to see if users would, in fact, organically find individuals to engage in conversation with.
Table \ref{table:average} shows that while matches with other users ranged from 6-10 per participant, the number of conversations that were initiated per user was less than 1. 

\begin{table}[!ht]
\begin{flushleft}
\setlength\extrarowheight{4pt}
\centering
\caption{Average user-to-user interactions at happening.}
\label{table:average}
\begin{tabular}{p{0.12\textwidth}|p{0.07\textwidth}|p{.12\textwidth}|p{.05\textwidth}}
\hline
\textbf{Count of} & \textbf{Average} & \textbf{Range (Min, Max)} & \textbf{SD} \\ 
\hline
Matches & 10.167 & (6, 11) & 1.403  \\ 
\hline
Friends & 2.667 & (1, 6) & 1.670  \\ 
\hline
Conversation Initiated & 0.917 & (0, 3) & 0.996 \\ 
\hline
Replies to Initiated Conversation & 0.583 & (0, 2) & 0.669 \\ 
\hline
Total Messages Sent & 2.75 & (0, 10) & 3.019 \\ 
\hline
\end{tabular}
\end{flushleft}
\end{table}

Through observations of the activity data, it is clear that the users in our happening did not have the same sense of purpose in finding someone to meet-up with as was outlined in survey scenarios.
Some users had difficulty utilizing the embedded chat features, while others were unsure how to best proceed with the conversation.
Conversations typically started with greetings and then, the new acquaintances started to coordinate a meet up. The purpose of the chat feature was intended to facilitate the meet-up and users were in a relatively close proximity to one another, and easily spotted from across a room. The chat was also used to debrief about the activity itself. We believe that the chat feature was instrumental to coordinating meeting up and created an opportunity for users to help one another familiarize themselves with the social aspects of the activity. Our observations indicate that the chat feature was not used in lieu of a face-to-face meeting at this event. 

\subsection{Survey Results of the Walk Through Event}
At the conclusion of the event, participants were asked to describe their experience in a survey.
Participants were asked to reflect on the experience overall in addition to answering questions about the potential community-centered activities that a user of this app would engage in.
Table \ref{table:afterHappening} shows that learning about others' perspectives and discovering new resources were rated higher than other activities, consistent with our survey, however, after use of the prototype, overall rankings were lower than those of potential users that viewed scenarios of fictitious characters alone.
While our hypothetical scenarios resulted in a character learning about others' perspective through a conversation or sharing a task, the users that attended our event did not experience these deeper shared conversations during the event.
These results indicate that further testing may be needed to understand hesitations to have conversations with other participants and tools that may be useful in facilitating citizens to connect with one another to discuss common interest areas and perform tasks.

\begin{table}[!ht]
\setlength\extrarowheight{4pt}
\centering
\caption{Responses related to community-centered activities (scale from 1-10).}
\label{table:afterHappening}
\begin{tabular}{m{15em}|m{2.5cm}}
\hline
\textbf{Survey Questions}\\Would you use this app to... & \textbf{After using app prototype}\\
\hline
...have a text-based conversation (through a text messages) rather than meet with someone in person? & 4.60\\
\hline
...meet new people? & 4.33\\ 
\hline
...learn about others perspectives in your community? & 5.00\\ 
\hline
...discover new resources in your community? & 5.50 \\ 
\hline
...discuss issues that I would NOT talk about otherwise? & 3.73 \\
\hline
...to fill downtime that might otherwise be spent  NOT being engaged in my community? & 3.82 \\
\hline
\end{tabular}
\end{table}

The survey also contained questions related to the application design.
In order to gain a better understanding of specific features in Community Animator application, we asked for open-ended feedback from the participants.
Users reported that they enjoyed the overall concept and idea of the app, that it seemed \textit{``like a dating app for finding friends."}
They also reported that the map view (rather than the list view) was a fast and easy way to find people that were close to them, but they were unsure if they would like their location visible to other users.
Instead, some users indicated that the list view of individuals in the same area, might protect their privacy to some degree.
All users reported that they would be willing to use the app in the future.  

\subsection{Focus Group with Members of Community Hub}
We have partnered with a local, government sponsored organization to distribute our application to its member population and examine its effects in enabling activities among currently unconnected segments of the local community. This organization is located in a space within the local municipal building and is a hub of collaboration among local organizations, ranging from emerging startups to established organizations. This event was announced in advance and a focus group was scheduled at the end of the three-hour period and posted to their website and digital calendars. 

During the three-hour usage period, members of our research team asked participants to download and use the Community Animator app to engage in new connections in the space and encouraged those who downloaded the app to attend the one-hour focus group.
During the event, 30 people used Community Animator, 17 people declined participation and 12 participants engaged in the focus group.
During the focus group, a moderator led a conversation about features and outcomes of interactions, such as meeting new people, learning about others perspectives, discovering new resources, engaging in new discussion topics, and filling downtown. 

Half of all participants (six individuals) in the focus group admitted to only creating an account, and not attempting to initiate an interaction after creating an account.
These participants stated that they had come to the space willing to participate, but became busy with other routine activities that they engage in regularly.
Although they did not review the app, they said they were interested in the idea and felt they would use Community Animator during downtime in the future. 

Of the six individuals that engaged with the app, four identified contacts with a shared interest area and two used it to reach out to a new contact and perform a task. Others said that they learned an acquaintance had a shared interest area that they were unaware of previously. Upon learning of this shared interest, they engaged in some conversation with the acquaintance about that shared interest. In a discussion about learning about new resources, the participants described human capital as the resource that they were most interested in connecting with.
One participant said: 

\begin{center}
\textit{``The value of [the community hub] is the opportunity it creates for me to find partners in my projects that serve the local community. This app could help me to identify partners for those projects, but I’m often looking for specific skill sets or availability."}
\end{center}

The desire to find others with similar skill sets was echoed by others in the focus group. The conversation turned to refinement of the tasks and interest area check boxes that appear when creating an account. Many participants would like to see specific local projects as task areas and areas for conversation. In addition, participants were also simply interested in the app as a tool for awareness by viewing profiles rather than engaging in meet-ups directly.

%% file: 50_Discussion.tex
\section{Discussion}
Through the design and implementation of this app, overall, we received positive outcomes that were consistent with our goal of promoting meet-ups based on interest-based similarities. Despite this step in the right direction, additional iterative design processes and testing will be necessary to further users toward our goal of using the app to achieve community-centered meet-ups. Although potential users rated interest and task based scenarios at a rating of 6-7 on the 10-point Likert scale, through hands-on use of the app, users rated its ability to meet these civic needs much lower. Part of the reason for these low responses was due to user frustrations with the interface; however, there may also be differences in perception when put in the position to kindle a face-to-face conversation, rather than watching a video that describes another person doing this activity. Future testing might explore the effect of the depth of conversations that users experience as it correlates to their perceived contribution of the app to civic life. 

While research questions 1 and 2 have been explored through the results, research question 3 requires additional analysis. The discrepancy between survey scenario-based response to survey question compared the use of the same survey questions to examine the prototype is an interesting result, and one that bears further exploration. Previous literature has designed tools to spark networking in situations where participants already have an interest and motivation in networking \cite{alt2011designing,heeks2007analyzing,McCarthy:2009:SCT:1556460.1556493,Memarovic:2012:DIP:2406367.2406420}, or seeding conversation for those that are already networking \cite{Xu:2014:SFC:2556420.2556789}. Future iterations may explore the use of this app in differing settings and situations to understand the need for these additional features. We believe further testing is necessary to understand the use of these types of app ``in the wild" when users are not specifically asked to turn it on and meet-up with others.

%% file: 60_Conclusion.tex
\section{Limitations \&\ Future Work}
A limitation of this work is the short duration of time that participants used Community Animator.
We are now working on a follow-up version of the application that can be used in a long-term implementation to further develop our analysis capabilities in order to pick up more cues to achieving the community-centered activities that we are interested in.
Also, we are working to discover ways to preserve users conversations in a low effort manner so that important community conversations can be shared with a wider audience.  

Another limitation of this app will be the identification of a critical mass. We see potential to test this app with local organizations that hire and train Community Animators. This can function as a check-in system for members in these spaces, which are beneficial to their work, and researchers can encourage adoption beyond these spaces. The activity traces in this app can be used to identify trending civic task categories within a community, hotspots of local conversation (such as third places), and map the increasing connectedness of community members with various organizational affiliations. Little research has been done to quantify the community impacts of increasing civic relationships on local organizations. This initial work contributes to that goal by identifying how these community interactions can create collaborations among divergent/siloed groups.

\section{Conclusion}
This work has identified community animation as a design space to leverage geosocial networking technology. We are interested in connecting communities as a whole, and this paper begins with the first step towards that end. The survey and evaluation that is presented in this paper is one piece of a larger project. Our study addresses two components: the first one we described is the likelihood and value of using the app, and the second examined the hypothesized contribution of various design scenarios. We explored in this study which of three scenarios is most likely to contribute to the hypothesized contribution. Our view of the survey results is not that users' preference for task-based meet-ups is a lapse into instrumentality, but rather as a more active expression of community-centric activities. Our review of community informatics literature shows that being informed (reading the newspaper) is sometimes a proxy for being engaged in the community. We believe that the results of the survey hold individual validity and importance and can be seen as a contribution from the development process.

Based on the survey a prototype that utilizes task-based meet-ups was designed. The Community Animator application uses the GPS in your smartphone to find ``animated" citizens near you that are interested in collaborating in the same civic tasks, but also supports civic interest-based conversations. Community Animator has been investigated through the implementation of the app and the results of that analysis have been analyzed. We explore the effectiveness of these contributions with an evaluation of the application. We might interpret our results to mean that people are interested in the app perhaps as a way to coordinate and engage in tasks over finding news or discussing interests. This is good news in that people may view Community Animator as a way of doing tasks together rather than learning about them like a news feed, which is a higher level of engagement. From a design perspective, we believe that the result of the testing supported our hypothesis that interest-based match would be most useful for users to engage in community-centered activities. 

Today, the modern casual encounter has more frequently become an activity coordinated by algorithms and geolocations \cite{pewresearch2013, smith2013online,Masden:2015:URC:2702123.2702417}, as many people have changed the way that they make friends and interact with others \cite{smith2013online,turkle2015reclaiming}. As society has become more accustomed to these behaviors, mobile applications that host introductions appear among ``most downloaded" lists in app stores and are widely discussed in popular culture \cite{elangovan2015factors}. The same technologies that have been used to inspire friendships and dating, can also be used within community technologies to reduce the social transaction costs of participation in civic networks \cite{carroll2006social,hampton2003neighboring,hampton2015ineighbor,hargittai2013digitally}. In such networks, inspiring interactions among citizens has the potential to lead to direct economic impacts for organizations and the local market \cite{campbell2013beyond,knight2010}.

%% file: proceedings.bbl

\begin{thebibliography}{00}


\ifx \showCODEN    \undefined \def \showCODEN     #1{\unskip}     \fi
\ifx \showDOI      \undefined \def \showDOI       #1{{\tt DOI:}\penalty0{#1}\ }
  \fi
\ifx \showISBNx    \undefined \def \showISBNx     #1{\unskip}     \fi
\ifx \showISBNxiii \undefined \def \showISBNxiii  #1{\unskip}     \fi
\ifx \showISSN     \undefined \def \showISSN      #1{\unskip}     \fi
\ifx \showLCCN     \undefined \def \showLCCN      #1{\unskip}     \fi
\ifx \shownote     \undefined \def \shownote      #1{#1}          \fi
\ifx \showarticletitle \undefined \def \showarticletitle #1{#1}   \fi
\ifx \showURL      \undefined \def \showURL       #1{#1}          \fi

\bibitem{animation2010}
 2010.
\newblock {Centre for Social Innovation} What is Community Animation?
\newblock \url{http://socialinnovation.ca/blog/what-community-animation}.
  (2010).
\newblock
\newblock
\shownote{Accessed: 2016-07-14.}


\bibitem{knight2010}
 2010.
\newblock {Knight Foundation} Knight Communities Overall.
\newblock \url{http://knightfoundation.org/sotc/overall-findings/}.   (2010).
\newblock
\newblock
\shownote{Accessed: 2014-07-14.}


\bibitem{pewresearch2013}
 2013.
\newblock {Pew Research Center} 5 facts about online dating.
\newblock \url{http://www.pewresearch.org/}.   (2013).
\newblock
\newblock
\shownote{Accessed: 2015-02-04.}


\bibitem{storyboard2014}
 2014.
\newblock {Story Board That} Digital StoryTelling. Powerful Visual
  Communication.
\newblock \url{ http://www.storyboardthat.com/}.   (2014).
\newblock
\newblock
\shownote{Accessed: 2014-09-16.}


\bibitem{amico2015}
 2015.
\newblock {Amico Bracelets}Bring Social Networks to Life with Style.
\newblock \url{http://www.amicobracelets.com/}.   (2015).
\newblock
\newblock
\shownote{Accessed: 2015-09-16.}


\bibitem{knightcivic2015}
 2015.
\newblock {D. Patel, M. Sotsky, J., Gourley, S., and Houghton. } The Emergence
  of Civic Tech: Investments in a growing field.
\newblock \url{http://knightfoundation.org/features/civictech/}.   (2015).
\newblock
\newblock
\shownote{Accessed: 2015-02-04.}


\bibitem{coalition2015}
 2015.
\newblock Edmonton Multicultural Coalition. Community Animators Recruitment.
\newblock
  \url{http://www.emcoalition.ca/wp-content/uploads/2012/01/Community-Animators.pdf}.
    (2015).
\newblock
\newblock
\shownote{Accessed: 2015-09-24.}


\bibitem{enacademic2015}
 2015.
\newblock Enacademic. Community Animator.
\newblock \url{http://new_words.enacademic.com/828/community_animator}.
  (2015).
\newblock
\newblock
\shownote{Accessed: 2015-09-24.}


\bibitem{hampton2015ineighbor}
 2015.
\newblock {Keith Hampton} i-Neighbors: free neighborhood website, email forum,
  social network for your community.
\newblock \url{https://www.i-neighbors.org/}.   (2015).
\newblock
\newblock
\shownote{Accessed: 2015-02-04.}


\bibitem{linkedin2015}
 2015.
\newblock {LinkedIn} Top 25 Community Animator profiles | LinkedIn.
\newblock \url{https://www.linkedin.com/title/community-animator}.   (2015).
\newblock
\newblock
\shownote{Accessed: 2015-09-24.}


\bibitem{outboxexperiment}
 2016.
\newblock OUTBOX - Think Outside The Office.
\newblock \url{http://www.downtownsilverspring.com/pages/silverspring-outbox}.
   (2016).
\newblock
\newblock
\shownote{Accessed: 2017-01-24.}


\bibitem{thirdplacesbuilders}
 2016.
\newblock ``Third Places" as Community builders.
\newblock
  \url{https://www.brookings.edu/blog/up-front/2016/09/14/third-places-as-community-builders/}.
    (September 2016).
\newblock
\newblock
\shownote{Accessed: 2017-01-24.}


\bibitem{alt2011designing}
{Florian Alt}, {Nemanja Memarovic}, {Ivan Elhart}, {Dominik Bial}, {Albrecht
  Schmidt}, {Marc Langheinrich}, {Gunnar Harboe}, {Elaine Huang}, {and}
  {Marcello~P Scipioni}. 2011.
\newblock \showarticletitle{Designing shared public display
  networks--implications from today’s paper-based notice areas}. In {\em
  International Conference on Pervasive Computing}. Springer, 258--275.
\newblock


\bibitem{campbell2013beyond}
{Tim Campbell}. 2013.
\newblock {\em Beyond smart cities: how cities network, learn and innovate}.
\newblock Routledge.
\newblock


\bibitem{carroll2006social}
{J Carroll}, {M Rosson}, {Andrea Kavanaugh}, {Daniel Dunlap}, {Wendy Schafer},
  {Jason Snook}, {and} {Philip Isenhour}. 2006.
\newblock \showarticletitle{Social and civic participation in a community
  network}.
\newblock {\em Computers, phones, and the Internet: Domesticating information
  technology\/} (2006), 168--181.
\newblock


\bibitem{carroll2013co}
{John~M Carroll}. 2013.
\newblock \showarticletitle{Co-production scenarios for mobile time banking}.
  In {\em International Symposium on End User Development}. Springer, 137--152.
\newblock


\bibitem{carroll2014grounding}
{John~M Carroll}, {Jessica Kropczynski}, {and} {Kyungsik Han}. 2014.
\newblock \showarticletitle{Grounding Activity in People-Centered Smart
  Territories by Enhancing Community Awareness.}
\newblock {\em IxD\&A\/}  {20} (2014), 9--22.
\newblock


\bibitem{carroll2003community}
{John~M Carroll} {and} {Debbie~Denise Reese}. 2003.
\newblock \showarticletitle{Community collective efficacy: Structure and
  consequences of perceived capacities in the Blacksburg Electronic Village}.
  In {\em System Sciences, 2003. Proceedings of the 36th Annual Hawaii
  International Conference on}. IEEE, 10--pp.
\newblock


\bibitem{Carroll:2005:CEM:1054972.1054974}
{John~M. Carroll}, {Mary~Beth Rosson}, {and} {Jingying Zhou}. 2005.
\newblock \showarticletitle{Collective Efficacy As a Measure of Community}. In
  {\em Proceedings of the SIGCHI Conference on Human Factors in Computing
  Systems} {\em (CHI '05)}. ACM, New York, NY, USA, 1--10.
\newblock
\showISBNx{1-58113-998-5}
\showDOI{%
\url{http://dx.doi.org/10.1145/1054972.1054974}}


\bibitem{carroll2014presence}
{John~M Carroll}, {Patrick~C Shih}, {Blaine Hoffman}, {Jing Wang}, {and}
  {Kyungsik Han}. 2014.
\newblock \showarticletitle{Presence and hyperpresence: implications for
  community awareness}.
\newblock {\em Interacting with presence: HCI and the sense of presence in
  computer-mediated environments. De Gruyter Open, Warsaw, Poland\/} (2014).
\newblock


\bibitem{carroll2015community}
{John~M Carroll}, {Patrick~C Shih}, {and} {Jessica Kropczynski}. 2015.
\newblock \showarticletitle{Community informatics as innovation in
  sociotechnical infrastructures}.
\newblock {\em The Journal of Community Informatics\/} {11}, 2 (2015).
\newblock


\bibitem{dimmick2010news}
{John Dimmick}, {John~Christian Feaster}, {and} {Gregory~J Hoplamazian}. 2010.
\newblock \showarticletitle{News in the interstices: The niches of mobile media
  in space and time}.
\newblock {\em New Media \& Society\/} (2010).
\newblock


\bibitem{elangovan2015factors}
{N Elangovan} {and} {Prachi Agarwal}. 2015.
\newblock \showarticletitle{Factors Influencing User Perception on Mobile
  Social Networking Apps}.
\newblock {\em Sumedha Journal of Management\/} {4}, 2 (2015), 27.
\newblock


\bibitem{gurstein2003effective}
{Michael Gurstein}. 2003.
\newblock \showarticletitle{Effective use: A community informatics strategy
  beyond the digital divide}.
\newblock {\em First Monday\/} {8}, 12 (2003).
\newblock


\bibitem{hampton2003neighboring}
{Keith Hampton} {and} {Barry Wellman}. 2003.
\newblock \showarticletitle{Neighboring in Netville: How the Internet supports
  community and social capital in a wired suburb}.
\newblock {\em City \& Community\/} {2}, 4 (2003), 277--311.
\newblock


\bibitem{han2015s}
{Kyungsik Han}, {Patrick~C Shih}, {Victoria Bellotti}, {and} {John~M Carroll}.
  2015.
\newblock \showarticletitle{It's Time There Was an App for That Too: A
  Usability Study of Mobile Timebanking}.
\newblock {\em International Journal of Mobile Human Computer Interaction
  (IJMHCI)\/} {7}, 2 (2015), 1--22.
\newblock


\bibitem{han2014local}
{Kyungsik Han}, {Patrick~C Shih}, {and} {John~M Carroll}. 2014.
\newblock \showarticletitle{Local news chatter: augmenting community news by
  aggregating hyperlocal microblog content in a tag cloud}.
\newblock {\em International Journal of Human-Computer Interaction\/} {30}, 12
  (2014), 1003--1014.
\newblock


\bibitem{han1916being}
{Kyungsik Han}, {Richard Wirth}, {Benjamin~V Hanrahan}, {Jiawei Chen}, {Sooyeon
  Lee}, {and} {John~M Carroll}. 2016.
\newblock \showarticletitle{Being connected to the local community through a
  Festival mobile application}.
\newblock {\em IConference 2016 Proceedings\/} (2016).
\newblock


\bibitem{Han:2014:SAM:2556420.2556824}
{Kyungsik~(Keith) Han}. 2014.
\newblock \showarticletitle{Studying the Application of Mobile Technology to
  Local Communities}. In {\em Proceedings of the Companion Publication of the
  17th ACM Conference on Computer Supported Cooperative Work \&\#38; Social
  Computing} {\em (CSCW Companion '14)}. ACM, New York, NY, USA, 61--64.
\newblock
\showISBNx{978-1-4503-2541-7}
\showDOI{%
\url{http://dx.doi.org/10.1145/2556420.2556824}}


\bibitem{hargittai2013digitally}
{Eszter Hargittai} {and} {Aaron Shaw}. 2013.
\newblock \showarticletitle{Digitally savvy citizenship: The role of internet
  skills and engagement in young adults' political participation around the
  2008 presidential election}.
\newblock {\em Journal of Broadcasting \& Electronic Media\/} {57}, 2 (2013),
  115--134.
\newblock


\bibitem{heeks2007analyzing}
{Richard Heeks} {and} {Savita Bailur}. 2007.
\newblock \showarticletitle{Analyzing e-government research: Perspectives,
  philosophies, theories, methods, and practice}.
\newblock {\em Government information quarterly\/} {24}, 2 (2007), 243--265.
\newblock


\bibitem{Kavanaugh:2013:ECP:2479724.2479750}
{Andrea Kavanaugh}, {Ankit Ahuja}, {Manuel P{\'e}rez-Qui\~{n}ones}, {John
  Tedesco}, {and} {Kumbirai Madondo}. 2013.
\newblock \showarticletitle{Encouraging Civic Participation Through Local News
  Aggregation}. In {\em Proceedings of the 14th Annual International Conference
  on Digital Government Research} {\em (dg.o '13)}. ACM, New York, NY, USA,
  172--179.
\newblock
\showISBNx{978-1-4503-2057-3}
\showDOI{%
\url{http://dx.doi.org/10.1145/2479724.2479750}}


\bibitem{Kavanaugh:2012:LNA:2307729.2307736}
{Andrea Kavanaugh}, {Samah Gad}, {Sloane Neidig}, {Manuel~A.
  P{\'e}rez-Qui\~{n}ones}, {John Tedesco}, {Ankit Ahuja}, {and} {Naren
  Ramakrishnan}. 2012.
\newblock \showarticletitle{(Hyper) Local News Aggregation: Designing for
  Social Affordances}. In {\em Proceedings of the 13th Annual International
  Conference on Digital Government Research} {\em (dg.o '12)}. ACM, New York,
  NY, USA, 30--39.
\newblock
\showISBNx{978-1-4503-1403-9}
\showDOI{%
\url{http://dx.doi.org/10.1145/2307729.2307736}}


\bibitem{Masden:2015:URC:2702123.2702417}
{Christina Masden} {and} {W.~Keith Edwards}. 2015.
\newblock \showarticletitle{Understanding the Role of Community in Online
  Dating}. In {\em Proceedings of the 33rd Annual ACM Conference on Human
  Factors in Computing Systems} {\em (CHI '15)}. ACM, New York, NY, USA,
  535--544.
\newblock
\showISBNx{978-1-4503-3145-6}
\showDOI{%
\url{http://dx.doi.org/10.1145/2702123.2702417}}


\bibitem{McCarthy:2009:SCT:1556460.1556493}
{Joseph~F. McCarthy}, {Shelly~D. Farnham}, {Yogi Patel}, {Sameer Ahuja},
  {Daniel Norman}, {William~R. Hazlewood}, {and} {Josh Lind}. 2009.
\newblock \showarticletitle{Supporting Community in Third Places with Situated
  Social Software}. In {\em Proceedings of the Fourth International Conference
  on Communities and Technologies} {\em (C\&\#38;T '09)}. ACM, New York, NY,
  USA, 225--234.
\newblock
\showISBNx{978-1-60558-713-4}
\showDOI{%
\url{http://dx.doi.org/10.1145/1556460.1556493}}


\bibitem{Memarovic:2012:DIP:2406367.2406420}
{Nemanja Memarovic}, {Marc Langheinrich}, {Elisa Rubegni}, {Andreia David},
  {and} {Ivan Elhart}. 2012.
\newblock \showarticletitle{Designing "Interacting Places" for a Student
  Community Using a Communicative Ecology Approach}. In {\em Proceedings of the
  11th International Conference on Mobile and Ubiquitous Multimedia} {\em (MUM
  '12)}. ACM, New York, NY, USA, Article 43, 10 pages.
\newblock
\showISBNx{978-1-4503-1815-0}
\showDOI{%
\url{http://dx.doi.org/10.1145/2406367.2406420}}


\bibitem{Nguyen:2015:KSS:2702123.2702411}
{Tien~T. Nguyen}, {Duyen~T. Nguyen}, {Shamsi~T. Iqbal}, {and} {Eyal Ofek}.
  2015.
\newblock \showarticletitle{The Known Stranger: Supporting Conversations
  Between Strangers with Personalized Topic Suggestions}. In {\em Proceedings
  of the 33rd Annual ACM Conference on Human Factors in Computing Systems} {\em
  (CHI '15)}. ACM, New York, NY, USA, 555--564.
\newblock
\showISBNx{978-1-4503-3145-6}
\showDOI{%
\url{http://dx.doi.org/10.1145/2702123.2702411}}


\bibitem{oldenburg1989great}
{Ray Oldenburg}. 1989.
\newblock {\em The great good place: Caf{\'e}, coffee shops, community centers,
  beauty parlors, general stores, bars, hangouts, and how they get you through
  the day}.
\newblock Paragon House Publishers.
\newblock


\bibitem{Rosson:2001:UES:501581}
{Mary~Beth Rosson} {and} {John~M. Carroll}. 2002.
\newblock {\em Usability Engineering: Scenario-based Development of
  Human-computer Interaction}.
\newblock Morgan Kaufmann Publishers Inc., San Francisco, CA, USA.
\newblock
\showISBNx{1-55860-712-9}


\bibitem{shih2015engaging}
{PC Shih}, {K Han}, {U Heo}, {and} {JM Carroll}. 2015.
\newblock \showarticletitle{Engaging Community Members with Digitally Curated
  Social Multimedia Content at an Arts Festival}.
\newblock {\em ACM Journal on Computing and Cultural Heritage, Under Review\/}
  (2015).
\newblock


\bibitem{smith2013online}
{Aaron Smith} {and} {Maeve Duggan}. 2013.
\newblock \showarticletitle{Online dating \& relationships}.
\newblock {\em Pew Internet \& American Life Project\/} (2013).
\newblock


\bibitem{turkle2015reclaiming}
{Sherry Turkle}. 2015.
\newblock {\em Reclaiming conversation: The power of talk in a digital age}.
\newblock Penguin Press HC.
\newblock


\bibitem{Xu:2014:SFC:2556420.2556789}
{Bin Xu}, {Tina Chien-Wen Yuan}, {Susan~R. Fussell}, {and} {Dan Cosley}. 2014.
\newblock \showarticletitle{SoBot: Facilitating Conversation Using Social Media
  Data and a Social Agent}. In {\em Proceedings of the Companion Publication of
  the 17th ACM Conference on Computer Supported Cooperative Work \&\#38; Social
  Computing} {\em (CSCW Companion '14)}. ACM, New York, NY, USA, 41--44.
\newblock
\showISBNx{978-1-4503-2541-7}
\showDOI{%
\url{http://dx.doi.org/10.1145/2556420.2556789}}


\end{thebibliography}
